# Small deviations from the $R^{1/4}$ law, the Fundamental Plane, and phase densities of elliptical galaxies

Jens Hjorth[1] and Jes Madsen

*Institute of Physics and Astronomy, University of Aarhus, DK-8000 Århus C, Denmark*

## ABSTRACT

The light profiles of elliptical (E) galaxies are known to display small systematic deviations (0.1–0.2 mag.) from the $R^{1/4}$ law. In this paper we show that the senses and amplitudes of these departures can be naturally accounted for by a simple distribution function constructed on the basis of statistical mechanics of violent relaxation. As a consequence, detailed light-profiles can be used to infer about the central potentials of E galaxies (the only free shape parameter of our model). Furthermore, the small deviations have recently been shown to correlate with luminosity, $\mathcal{L}$. This observation entails a slight breaking of the generally assumed structural homology between E galaxies. Using our model we parametrize this broken homology by establishing a correlation between a suitably normalized central potential and luminosity. The non-homology means that a basic assumption in the interpretation of the Fundamental Plane (FP) breaks down, and with it the conclusions derived from it. Instead, by assuming that $\mathcal{M}/\mathcal{L}$ is independent of luminosity, we derive a relation akin to the FP directly from our correlation. This implies that the FP may have a simple stellar dynamical origin. We can reproduce the observed Carlberg–Kormendy relation for the central phase-space densities ($f_c$) of E galaxies of identical structure, $f_c \propto \mathcal{L}^{-2.35}$, but non-homology changes it to a much weaker dependence ($f_c \propto \mathcal{L}^{-1.5}$ for constant $\mathcal{M}/\mathcal{L}$) which would imply that the central phase-space densities of ellipticals are comparable to those of spiral galaxies ($\lesssim 10^{-6} \mathrm{M}_\odot \mathrm{pc}^{-3} (\mathrm{km\,s}^{-1})^{-3}$). Thus dissipationless merging is consistent with the FP, although *HST* observations, notably the presence of nuclear embedded disks, indicate that the assumptions behind our model (isotropy and pressure support) break down for the nuclei of faint E galaxies in which dissipative processes seem to be important.

*Subject headings:* celestial mechanics, stellar dynamics — galaxies: elliptical and lenticular, cD — galaxies: fundamental parameters — galaxies: kinematics and dynamics — galaxies: photometry — galaxies: structure

---

[1] Present address: Institute of Astronomy, Madingley Road, Cambridge CB3 0HA, UK; jens@mail.ast.cam.ac.uk

## 1. Introduction

Normal, bright elliptical (E) galaxies follow the empirical $R^{1/4}$ law (de Vaucouleurs 1948),

$$\mu(R) = \mu_e + 8.3268\left[\left(\frac{R}{R_e}\right)^{1/4} - 1\right], \quad (1)$$

which is a linear relation between the surface brightness, $\mu$, and the fourth root of the projected radius, $R$. The effective radius, $R_e$, is defined as the radius of the isophote containing half the total light, and $\mu_e = \mu(R_e)$ is the corresponding surface brightness.

In itself the universality of this surface-brightness profile is a remarkable property, especially in view of the very different sizes, fine structures, and formation histories of E galaxies. But these systems follow additional tight relations.

The Fundamental Plane (FP) (Djorgovski & Davis 1987; Dressler et al. 1987) confines the manifold of E galaxies to a two-dimensional sheet in the three-dimensional space of characteristic luminosity, radius, and velocity dispersion. This makes E galaxies a two-parameter family. The version of the FP to be used in this paper takes the form

$$\mathcal{L} \propto R_e^A \sigma^{2B}; \quad A \approx B \approx 0.8 \pm 0.1 \quad (2)$$

(e.g., Schaeffer et al. 1993). This relation is believed to be a consequence of the virial theorem, the homology between elliptical galaxies (the $R^{1/4}$ law), and a weak dependence of the mass-to-light ratio, $\mathcal{M}/\mathcal{L}$, on $\mathcal{L}$,

$$\mathcal{M}/\mathcal{L} \propto \mathcal{L}^\alpha; \quad 1/\alpha \approx 6 \pm 2 \quad (3)$$

(Kormendy & Djorgovski 1989; Djorgovski & Santiago 1993). This could be due to dark matter, varying IMF or metallicity, etc. (Renzini & Ciotti 1993; Guzmán, Lucey & Bower 1993), but no convincing theoretical explanation or interpretation has been given. For example, many explanations need fine-tuning and are expected to have difficulties with merger processes which tend to mess up nice correlations. An intriguing observation in this respect is that the FP seems to persist at IR wavelengths where values of $\alpha \approx 0.1$–$0.15$ are found (Djorgovski & Santiago 1993; Recillas-Cruz et al. 1991).

The situation is slightly more satisfactory with respect to understanding the $R^{1/4}$ law. Although the physics of the violent relaxation process (Lynden-Bell 1967), which is believed to be an important ingredient in E galaxy formation, is still largely unresolved, the concept can be used in conjunction with statistical mechanics to give a reasonable explanation of the $R^{1/4}$ law (Hjorth & Madsen 1991, 1993, and references therein).

Ellipticals are known to exhibit small systematic deviations from the $R^{1/4}$ law, correlated with luminosity (Michard 1985; Schombert 1986). For example, Caon, Capaccioli & D'Onofrio (1993) have shown that fitting E galaxies with generalized "$R^{1/m}$ laws" leads to a correlation between $m$ and total luminosity. This means that the structural homology generally assumed for the class of E galaxies is slightly broken. Likewise, the end-products of numerical simulations frequently show characteristic residuals from the $R^{1/4}$ law (Londrillo, Messina & Stiavelli 1991).

Recently Burkert (1993) has treated CCD data of a large sample of elliptical galaxies (Bender, Döbereiner & Möllenhoff 1988; Franx, Illingworth & Heckman 1989; Peletier et al. 1990) in a systematic way. The combination of a sample of high-quality data with a consistent data analysis allows new insight into the issues mentioned above. The present paper is devoted to such an investigation. Our aim is to show that the observed departures from the $R^{1/4}$ law can be understood theoretically essentially as the signature of the isothermal sphere. Furthermore, we shall investigate to what extent the detected non-homology between E galaxies affects the usual interpretation of the FP. In fact, we shall argue that the FP may arise as a result of this slight breaking of the structural homology between E galaxies that nevertheless keep approximate $R^{1/4}$ law profiles. Finally, in order to discuss how mergers may affect the FP we show that our approach allows a new assessment of the central phase-space densities in E galaxies to the extent that these can be considered isothermal and pressure supported (e.g., do not harbor embedded disks).

The paper is built up as follows: In §2 we briefly review the approach and results of Burkert (1993) and introduce a model of elliptical galaxies based on the theory of violent relaxation (Hjorth & Madsen 1991). The message of this section is that the model is shown to be capable of reproducing the observed senses and amplitudes of the departures of the $R^{1/4}$ law, thereby giving them a possible physical explanation. Given that this explanation entails a slightly varying structure as a function of total luminosity we also describe two applications within this framework. In §3 we show that the Fundamental Plane can be understood as a result of these slightly dissimilar structures of E galaxies. Finally we compute the central phase-space densities of elliptical galaxies in § 4 and show that the luminosity-dependent structure of E galaxies has interesting implications for the hypothesis that E galaxies may form from dissipationless merging of spiral galaxies. In § 5 we discuss the evidence for a possible dichotomy between bright and faint galaxies in the light of findings of this paper. Conclusions are provided in § 6.

## 2. Small Deviations from the $R^{1/4}$ Law

### 2.1. Burkert's Analysis

We first briefly summarize the results of Burkert (1993) (for details, the reader is referred to the original paper). The radius coordinate is defined by taking the ellipticity of the observed galaxies properly into account. An effective "effective radius", $x_e$, is determined from the light profile in a self-consistent way: From a first estimate of $x_e$ (in arcsec) the best-fitting straight line is fitted to the data in the range $0.6 \leq x \leq 1.1$, where $x$ is the radius (in units of $x_e$) to the power of $1/4$. The lower limit in $x$ ensures that seeing effects are minimized and the upper limit is imposed to avoid flat fielding errors, sky subtraction errors, etc. A new guess of $x_e$ is then determined from $(8.3268/b)^4 x_e$, where $b$ is the best-fitting slope (see equation (1)). With this updated value of $x_e$ the procedure is repeated until convergence (typically 5 times).

63 bright E galaxies were studied in this way. 70 % of these light profiles showed a characteristic dip at $x \approx 0.8$ being fainter by typically 0.1 mag than the best-fitting $R^{1/4}$ law (for representative observed profiles we refer to Burkert's (1993) paper). The remaining galaxies typically had a maximum at $x \approx 0.9$. The sense and strength of these departures were quantified in the parameter

$$\delta b = \frac{1}{8.3268} \left( \left.\frac{\partial \mu}{\partial x}\right|_{x \in [x_{\rm cut}; 1.1]} - \left.\frac{\partial \mu}{\partial x}\right|_{x \in [0.6; x_{\rm cut}]} \right), \quad (4)$$

$x_{\rm cut}$ being defined by the minimum (maximum) of the bump at $x_{\rm cut} = 0.8\ (0.9)$. The characteristic minimum found in most galaxies gives a negative value of $\delta b$. In this way Burkert found that luminous galaxies have preferentially negative $\delta b$ and faint galaxies positive $\delta b$ (see also Fig. 3 below). Thus there is a systematic trend in the deviations from the $R^{1/4}$ law that is quantified in a parameter that can be computed consistently for various models.

### 2.2. Model

In this paper we choose to compare the observed profiles with those computed from a theoretical model constructed to account for the surface-brightness profiles of spherical galaxies.

Based on statistical mechanics and a simple violent relaxation scenario we have proposed a model for the phase-space density in elliptical galaxies (Hjorth & Madsen 1991, 1993). It involves a simple two-step scenario: First violent relaxation redistributes stars in phase space so efficiently that the original prediction by Lynden-Bell (1967) is applicable. However, this only takes place in a finite volume. The second step, therefore, is an escape of those particles that have gained positive energies during the first violent relaxation phase, and redistribution of weakly bound particles to larger average radii.

This scenario leads to a Maxwell–Boltzmann distribution in the central parts of the galaxy, $f \propto \exp(-E/\sigma^2)$; in the outer parts the differential energy distribution, i.e., the number of particles with a given energy per unit mass, $E$, (Binney 1982) is predicted to be finite, $N(E) \approx N_0$, for $E \to 0^-$ but zero above the escape energy (for details see Hjorth & Madsen (1991)). For numerical confirmations of these hypotheses see e.g. Londrillo et al. (1991).

We have previously discussed the properties of the model in terms of its one free shape parameter, namely the dimensionless central potential $\Psi_0 \equiv -\Phi_0/\sigma^2$. Here $\Phi_0$ is the physical central potential and $\sigma$ is a model parameter which in the interesting range of $\Psi_0$ is approximately equal to the one-dimensional central velocity dispersion which is again roughly equal to the line-of-sight velocity dispersion derived from observations. In these units the model predicts

$$\rho(\Psi) \propto \begin{cases} \exp(\Psi) & \text{in the inner regions;} \\ \Psi^n & \text{in the outer regions,} \end{cases} \quad (5)$$

where $n = 4.1 - 4.4$, the boundary being approximately at the effective radius. In the outer Keplerian regions, where $\Phi(r) \sim -r^{-1}$ this corresponds to a fall-off in density as $\rho \sim r^{-n}$. Recently, each of these predictions have been shown to be in perfect agreement with independent numerical $N$-body simulations of merger remnants (Lima-Neto 1994).

The model has been successful in retrieving the $R^{1/4}$ law for any $\Psi_0 \gtrsim 8.5$, has given a successful fit to the small departures from the $R^{1/4}$ law in the light curve of NGC 3379 (Hjorth & Madsen 1991; Sodemann & Thomsen 1994), and has been used to understand radial-orbit instability thresholds in dissipationless-collapse simulations (Hjorth 1994). In this Section we show that it can also reproduce the observed deviations from the $R^{1/4}$ law found by Burkert.

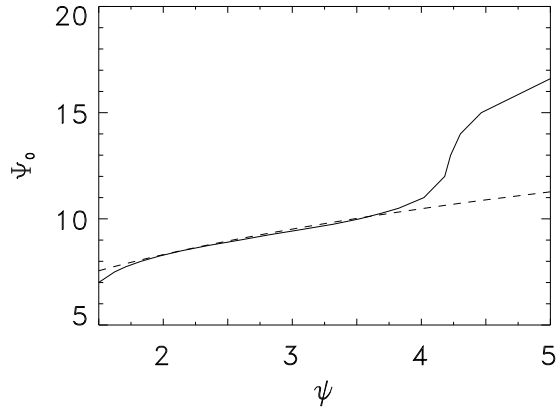

Fig. 1.— The $\Psi_0$ - $\psi$ relation. The dashed curve is the relation $\Psi_0 = 6.60\psi^{1/3}$ used in §2.2.

In the discussion that follows we shall make use of a more model-independent dimensionless central potential,

$$\psi \equiv -\frac{r_h}{G\mathcal{M}}\Phi_0 = \frac{r_h}{G\mathcal{M}}\Psi_0\sigma^2, \qquad (6)$$

where $r_h$ is the half-mass radius. For example, $\psi$ is readily computed in numerical $N$-body experiments (Londrillo et al. 1991). The unique relation between $\Psi_0$ and $\psi$ is shown in Figure 1.

This paper mainly deals with the central properties of the model (i.e., out to one or two effective radii). According to equation (5) this is essentially the isothermal sphere. This assertion is confirmed by computing the analogous curve to Figure 1 for the King model. The latter model also has an isothermal core, but a different cut-off in $f(E)$. The corresponding curve shows very much the same trend. Thus, although our discussion will be based on a specific, albeit physically well-motivated, model, many of our results do not hinge on this choice and our conclusion may be generally valid for most realistic models with a non-singular isothermal core.

Some useful expressions for the central properties derived from the model (to be used in §3 and §4) follow below.

The relation between the half-mass radius and the effective radius is $r_h/R_e \approx 1.34 \pm 0.01$ as for any spherical system approximately obeying the $R^{1/4}$ law (Young 1976). Using $\Psi_0 = 6.60\psi^{1/3}$ (cf. Fig. 1) valid for our model in the relevant range of values of $\psi$ (see §2.3) we obtain from equation (6) the expression

$$\sigma^2 \approx 0.113\frac{G\mathcal{M}}{R_e}\psi^{2/3}; \qquad 1.6 \lesssim \psi \lesssim 4, \qquad (7)$$

for the central one-dimensional velocity dispersion, which is correct to 1 % for $1.85 \leq \psi \leq 3.5$ and correct to 5 % for $1.6 \leq \psi \leq 1.85$ and $3.7 \leq \psi \leq 4.0$.

A useful expression for the central density is found to be

$$\rho_c \approx 0.0277\frac{\mathcal{M}}{R_e^3}10^\psi \quad (\pm 18\,\%); \qquad 1.6 \lesssim \psi \lesssim 4, \qquad (8)$$

and the central phase-space density, as defined by the model, is found to be well-approximated by

$$f_c \approx 0.064\rho_c\sigma^{-3} \quad (\pm 1\,\%), \qquad (9)$$

(consistent with an isothermal core, $f_c = (2\pi)^{-3/2}\rho_c\sigma^{-3}$), where central velocity dispersions and densities are calculated using equations (7) and (8).

### 2.3. Results

It is evident from Burkert's (1993) data that the iteratively determined $x_e$ does not give the "correct" half-light radius. Repeating the iterative determination of $x_e$ for our models as a function of $\psi$ confirms that $x_e$ may be up to a factor of 2 wrong. However, in the present context it is more important that the procedure is well-defined: Burkert's observational results can be directly compared with model predictions. Typical residuals from the $R^{1/4}$ law ($\mu_{\rm model} - \mu_{R^{1/4}}$) (using $x_e$, not $R_e$ as the effective radius) for various values of $\psi$ are shown in Figure 2.

The lower limit $\Psi_0 \gtrsim 8.5$ mentioned in §2.2, arising from the requirement that the model should give reasonable $R^{1/4}$ laws, leads to a lower limit on $\psi \gtrsim 2$. Less good but useful $R^{1/4}$ law approximations are obtained down to $\psi \approx 1.6$ but for these values the iterative determination of $x_e$ does not converge to a reasonable value. For these profiles only one redetermination of the slope was done. As a consequence, high positive values of $\delta b$ ($\gtrsim 0.3$) are uncertain. For such small values of $\psi$ a maximum around $x \approx 0.85 \pm 0.1$ is recovered leading to a positive $\delta b$.

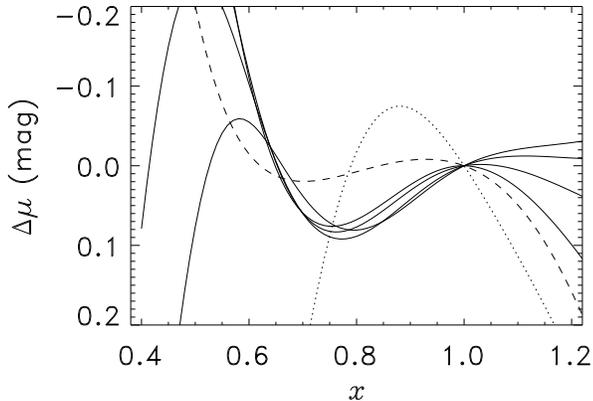

Fig. 2.— Computed departures from the $R^{1/4}$ law for $\psi = 1.71$ (dotted curve), $\psi = 2.59, 3.07, 3.52, 4.02$ (solid curves, from top to bottom at $x = 1.2$), and $\psi = 4.47$ (dashed curve).

For slightly larger values of $\psi$ $\delta b$ approaches 0. Then, for a large range of values of $\psi$ (from 2.5 to 4) the observed characteristic minimum at $x \approx 0.8$ is recovered (cf. §2.1) leading to a negative $\delta b$. The amplitude of the dip also corresponds to those observed in E galaxies. Thus, the model describes the small deviations found by Burkert satisfactorily.

For $\psi \gtrsim 4$ ($\Psi_0 \gtrsim 11$) this dip becomes less pronounced and $\delta b$ again approaches 0. Furthermore, the structure of the model changes dramatically. From Figure 1 it is evident that $\Psi_0$ and thus the central density of the model (cf. equation (5)) increases drastically above $\psi \approx 4$. A similar behavior is seen in, e.g., the King model (see also Fig. 4-10 in Binney & Tremaine (1987)). This is probably an upper limit related to the observed upper cut-off in luminosity and velocity dispersion for E galaxies.

This limit is furthermore consistent with numerical simulations: Londrillo et al. (1991) found in their cold dissipationless collapse simulations the deepest central potential to be $\psi = 3.20$, but dissipation can increase the density compared to this value somewhat (the simulation leading to the shallowest potential had $\psi = 1.45$ but gave a poor fit to the $R^{1/4}$ law). For the exact $R^{1/4}$ law $\psi = 3.54$ (Young 1976). In this connection it is interesting to note that from the published residuals from the $R^{1/4}$ law of Londrillo et al. (1991) it is evident that systems with $2 \lesssim \psi \lesssim 2.4$ have $\delta b \approx 0$ whereas $\delta b$ is negative for $2.4 \lesssim \psi \lesssim 3.2$, consistent with our analysis.

To summarize, for $2.5 < \psi < 4$ the model reproduces the decrease in brightness at $x \approx 0.8$ relative to the $R^{1/4}$ law observed in 70 % of bright E galaxies ($\delta b < 0$). For $1.6 < \psi < 2$ it can also reproduce the observed increased brightness relative to the $R^{1/4}$ law ($\delta b > 0$) for the remaining galaxies.

## 3. The Fundamental Plane

The question now is whether $\psi$ can be determined from photometry of elliptical galaxies. From the model we can calculate $\delta b$ as a function of $\psi$ and Burkert found a dependence of $\delta b$ on $\mathcal{L}$. To examine whether there is a relation between $\psi$ and $\mathcal{L}$ we now incorporate the FP in the discussion.

For the purposes of our discussion, the FP relation (2) can be written in the simplified form

$$\mathcal{L} \propto (R_e \sigma^2)^\gamma; \qquad 0.7 \lesssim \gamma \lesssim 0.9, \qquad (10)$$

without any decisive loss of generality. Combining the observational FP relation (10) with equation (7) we obtain

$$\psi^{2/3} \propto \frac{\mathcal{L}}{\mathcal{M}} \mathcal{L}^\alpha; \qquad \alpha = \frac{1}{\gamma} - 1. \qquad (11)$$

Two simple interpretations of this equation are possible:

(i) If $\psi$ is independent of $\mathcal{L}$ then $\mathcal{M}/\mathcal{L} \propto \mathcal{L}^\alpha$.

(ii) If $\mathcal{M}/\mathcal{L}$ is independent of $\mathcal{L}$ then $\psi^{2/3} \propto \mathcal{L}^\alpha \propto \mathcal{M}^\alpha$.

Normally structural homology (i) is assumed, i.e., that all galaxies have similar structures. In this case equation (3) is recovered. This is a reasonable hypothesis given that most E galaxies are well-described by an exact $R^{1/4}$ law.

The second possibility is to take the systematic departures from the $R^{1/4}$ law as an indication of dissimilar structures, corresponding to a $\psi$–$\mathcal{L}$ correlation. In the simplest case (ii) $\mathcal{M}/\mathcal{L}$ is independent of $\mathcal{L}$. This hypothesis can be tested by comparing the resulting value of $\alpha$ with that determined observationally.

The structure–luminosity trend can be visualized in a $\delta b$–$M_B$ diagram. Burkert's (1993) relation (his Fig. 8) is not unique, but this is not surprising given that the observations are difficult and that much of the fine structure of elliptical galaxies (e.g., disks, shells, dust, environment etc.) will undoubtedly spoil such a relation somewhat. What is interesting is that there does seem to be a trend: For $M_B \gtrsim -19.5$ ($M_B$ is the total absolute blue magnitude) $\delta b$ is positive whereas $\delta b < 0$ for all galaxies with $M_B \lesssim -22$. Furthermore, there seems to be a lower boundary on $\delta b$. These are the trends that we can hope to reproduce.

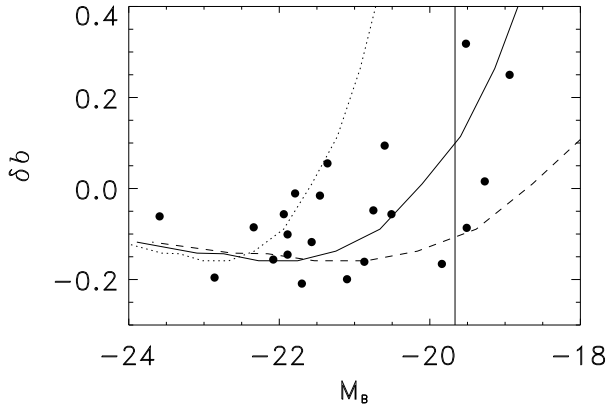

Fig. 3.— Circles are $\delta b$ values for a subsample of galaxies (Burkert 1993) with $C4 \leq 0$ and $\epsilon \leq 0.3$ containing no dust as a function of total $B$ magnitude, $M_B$. The thin vertical line is the approximate boundary between galaxies with $m > 4$ (left) and $m < 4$ (right) found by Caon et al. (1993). The curves are relations obtained using $M_B = -2.5 \log \psi^{2/(3\alpha)} - K$ with $(\alpha = 0.09, K = 12.5)$ (dotted curve), $(\alpha = 0.12, K = 15.5)$ (solid curve), $(\alpha = 0.18, K = 18.5)$ (dashed curve). All values of $M_B$ are scaled to $h = 0.5$.

In order to minimize the potential effects arising from possible contributions from disks in the E galaxies a subset of Burkert's sample is plotted in Figure 3. These data points only include galaxies with "boxiness parameter" $C4 \leq 0$ (e.g., Peletier et al. 1990) and ellipticity $\epsilon \leq 0.3$. Furthermore, only galaxies showing no signs of dust or other peculiarities are plotted. It is seen that there is still a clear trend although there are much fewer data points (22).

In a related study Caon et al. (1993) found that fitting $R^{1/m}$ laws in general leads to $m > 4$ for $M_B \lesssim -19.3$ and $m < 4$ for fainter galaxies. Similarly, Young & Currie (1994) have found that dwarf elliptical galaxies all have $m < 4$ with a conspicuous correlation between $m$ and $M_B$ extending the findings of Caon et al. (1993). Although its physical meaning is less clear, the parameter $m$ is similar to $\delta b$ in that higher $m$ leads to higher $\psi$ (Ciotti 1991) and that $m - 4$ and $\delta b$ have opposite signs. The findings of Caon et al. (1993) and Young & Currie (1994), which are based on a larger sample of galaxies covering a larger range in absolute luminosity, thus strengthen the conclusions derived from Figure 3.

Assuming that the possible range of $\psi$ for bright E galaxies suitably described by the model is $1.6 \lesssim \psi \lesssim 4$, the value of the exponent $\alpha$ (cf. equation (11)) is quite well-determined if the allowed range in $\psi$ must span the observed magnitude range. In Figure 3 we show a good fit to the $\delta b$ vs. $M_B$ data obtained by letting $\alpha = 0.12^{+0.06}_{-0.03}$ ($\gamma = 0.88 \pm 0.03$) under the hypothesis (ii) of constant $\mathcal{M}/\mathcal{L}_B$. This value of $\alpha$ is in agreement with observations, especially in the infrared ($\sim$bolometric) where Djorgovski & Santiago (1993) found $\alpha \approx 0.13$ and Recillas-Cruz et al. (1991) found $\alpha \approx 0.15$ for galaxies in Virgo and $\alpha \approx 0.09$ in Coma. Observations in infrared light are preferable since they more directly probe the old relaxed stellar population and have the advantage that line-blanketing effects are negligible. The observed larger values of $\alpha$ at shorter wavelengths is to a great deal probably due to a metallicity effect (Renzini & Ciotti 1993).

Assuming a correlation between the structure of galaxies (parametrized by $\psi$) and their luminosity we have thus shown that the model leads to a FP-like relation consistent with observations even with a constant mass-to-light ratio.

## 4. Central Phase-Space Densities

To look into the possible origin of this FP relation, in particular the effects of merging, we next compute the central phase-space densities of E galaxies. Dissipationless merging cannot increase the maximum (central) phase-space density (Tremaine, Hénon & Lynden-Bell 1986). Therefore, accurate determination of central phase-space densities may have important consequences for formation theories of elliptical galaxies, in particular the merger hypothesis (Vedel & Sommer-Larsen 1990).

Unfortunately, determination of central phase-space densities of elliptical galaxies is difficult. Previous determinations have used the core-fitting method (Richstone & Tremaine 1986) which is based on a fit of an isothermal model to the central parts of the galaxy. However, ground-based observations are strongly affected by seeing and most galaxies are known to have non-isothermal cores, with a significant excess of light near the center (Kormendy 1985a; Møller, Stiavelli & Zeilinger 1993). For example, recent *HST* (Hubble Space Telescope) data of the cores of ellipticals indicate very high central values, presumably due to central black holes (Crane et al. 1993).

We here propose a different method based on the indirect determination of the central potential using the light-profiles in the regions un-affected by seeing and the central cusp. In the adiabatic model the phase-space density is unaltered by a black hole so our assumption of an isothermal core may be adequate for the present purposes.

To compute physical phase-space densities (in units of $M_\odot \, pc^{-3} (km\,s^{-1})^{-3}$) using the expression for $f_c$ given in equation (9) we shall resort to an approximate relation between $R_e$ (in parsec) and $M_B$. For purposes of illustration we choose to combine the Hamabe–Kormendy (1977) relation $\mu_{eB} = 2.94 \log R_e + 11.93$ (valid for a distance modulus for the Virgo cluster of $(m-M) = 31.3$ and a Hubble constant $h = 0.7$; $H_0 = 100 h \, km\,s^{-1}\,Mpc^{-1}$, cf. Capaccioli, Caon & D'Onofrio (1992)) with $M_B = -5 \log R_e + \mu_{eB} - 24.05$ which follows from the $R^{1/4}$ law (Young 1976), to get $M_B = -2.06 \log R_e - 12.12 + 2.94 \log(0.7/h)$, i.e., $R_e \propto \mathcal{L}_B^{1.213}$. Equations (7) and (8) then lead to

$$f_c \propto 10^\psi \psi^{-1} \mathcal{L}_B^{-2.32}. \qquad (12)$$

This relation clearly illustrates that there is a dramatic effect of galaxies having $\psi$ dependent on luminosity: For the usual assumption that all ellipticals have a fixed structure ($\psi$ constant) one recovers (up to a prefactor depending of the choice of $\psi$) the relation of Carlberg (1986) who found $f_c \propto \mathcal{L}_B^{-2.35}$, using data of Kormendy (1985b). However, using our $M_B - \psi$ relation with $\alpha = 0.12$ a much weaker dependence of luminosity, $f_c \propto \mathcal{L}_B^{-1.5}$, is obtained, and the values are found to be quite low, $\lesssim 10^{-6} M_\odot \, pc^{-3} (km\,s^{-1})^{-3}$. These effects are illustrated in Figure 4.

Before discussing this result a word of caution is appropriate. The results shown in Figure 4 were deduced assuming a scaling relation between $R_e$ and $M_B$. The existence of the FP and the distribution of galaxies *on* the FP means that such a relation is only approximate and contains a lot of real scatter. What our discussion does illustrate, however, are the potentially important effects that the varying structure of ellipticals may have. Furthermore, assuming different relations between $R_e$ and $\mathcal{L}_B$ than the one adopted here results in different $f_c$–$M_B$ relations (cf. Fig. 4)

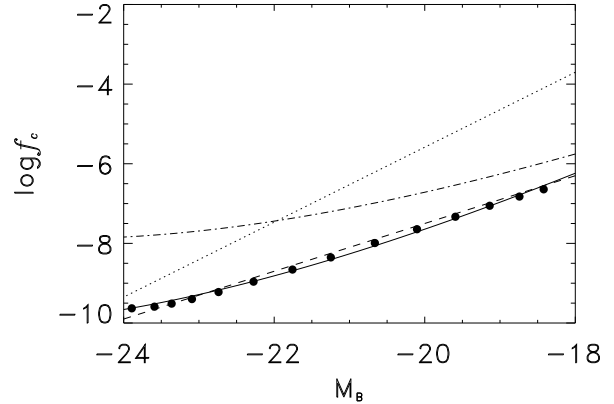

Fig. 4.— Central phase-space density $\log f_c$ (in units of $M_\odot \, pc^{-3} (km\,s^{-1})^{-3}$) versus $M_B$. Filled circles are exact values for the model using $R_e \propto \mathcal{L}_B^{1.213}$ (see text). The solid curve is calculated using the approximations given in the paper. The dashed line is the approximate relation $\log f_c \approx 0.6 M_B + 4.5$. The dotted line is the much steeper Carlberg–Kormendy relation (Carlberg 1986) valid if structural homology is assumed. The dashed-dotted curve is similar to the solid curve but has been derived using $R_e \propto \mathcal{L}_V^{0.85}$ (Guzmán et al. 1993) and $B - V = 0.95$. $\mathcal{M}/\mathcal{L}_B = 10$ and $h = 0.5$ have been assumed in this plot.

but the qualitative effects of the varying structure are unaltered.

## 5. Discussion

Taken at face value, the results that we find indicate that most ellipticals may have central phase-space densities that are comparable to those of spirals for which $f_c \approx (10^{-5}\text{--}10^{-7}) M_\odot \, pc^{-3} (km\,s^{-1})^{-3}$, cf. Carlberg (1986) and Lake (1989). Notice in particular that the central phase-space densities of faint elliptical galaxies are predicted to be about two orders of magnitude smaller than predicted by the Carlberg–Kormendy relation. The reason why the small deviations, i.e., the central potential, has such a dramatic influence on the measurement of $f_c$ can be traced back to the fact that any model with a sufficiently deep central potential can lead to the $R^{1/4}$ law. Assuming a unique profile for all galaxies corresponds to chosing a specific central potential. As we have shown here, this may be erroneous and can lead to substantially different derived results for the central properties of the galaxies.

As an illustration of this effect consider the data sample of Kormendy (1985b), upon which the Carlberg–Kormendy relation is based. The central projected surface brightnesses (and thus the central phase-space densities) of the galaxies were obtained by performing a seeing deconvolution of the observed light profiles. Following Schweizer (1981) it was assumed that the cores of the galaxies were described by a King model with $c = 2.25$ (corresponding to $\psi \approx 9.5$, or $M_B = -22.5$ for $\alpha = 0.12$, i.e., a quite deep central potential). But as Møller et al. (1993) have demonstrated the assumptions underlying the seeing deconvolution are very model dependent and may give misleading results.

It can be seen from Figure 4 that the Carlberg–Kormendy relation and the one found in this paper (equation (12)) agree quite well for the bright galaxies. These galaxies have the smallest seeing corrections (Kormendy 1985b) and are well-described by a $c = 2.25$ King model because they have deep dimensionless central potentials, $\psi$. The large discrepancies occur for the faint unresolved galaxies for which the seeing corrections (and thus the systematic uncertainties) are largest. Furthermore, according to the findings of the present paper, the faint galaxies have cores that are better described by King models with considerably smaller values for the concentration (down to $c \approx 1.6$). The seeing deconvolution would therefore overestimate the value of the central surface brightness and thereby the central phase-space density for the faint galaxies.

Still, the fact that our relation falls significantly below the Carlberg–Kormendy relation is somewhat disturbing. Although difficult to compute without a useful model for the distribution function recent *HST* observations (Crane et al. 1993) indicate that the central phase-space densities are indeed much higher than predicted by Figure 4, especially for faint galaxies. However, our model is only valid for E galaxies that are pressure supported and isothermal and there seems to be a dichotomy between bright and faint galaxies, the dividing line being approximately as indicated in Figure 3: Bright galaxies in general are resolved with the *HST* (Kormendy et al. 1994), do not contain nuclear disks and have boxy isophotes (Nieto, Bender & Surma 1991), belong to the 'bright group' of Capaccioli et al. (1992), are pressure supported (Davies et al. 1983), have $\delta b < 0$ and $m > 4$, and have central phase-space densities less than those for spirals. On the other hand, the faint E galaxies in general are unresolved with *HST*, they often contain nuclear dust and gaseous/stellar disks, sometimes with signs of star formation, belong to the 'ordinary group', are rotationally supported, have $\delta b > 0$ and $m < 4$, and have phase-space densities larger than those of spirals according to the Carlberg–Kormendy relation.

It is therefore possible that the discrepancy between the inferred high central phase-space densities from *HST* and our estimates in Figure 4 is a measure of the amount of dissipation that has occurred, i.e., bright galaxies may have formed essentially dissipationlessly through a merging hierarchy whereas (the nuclei of) faint E galaxies involved significant dissipation. Star formation and material in disks may significantly raise the inferred central phase-space density. It should be noted, however, that the 'underlying E galaxy' may well have a much lower central phase-space density and be well-described by our model further out (where $\delta b$ is measured). Also, the central properties affected by dissipational processes may be unimportant for the gross dynamical evolution of an E galaxy (excluding the nucleus) due to the very little mass fraction involved. In this case the dissipationless scenario may still be valid.

Thus, to the extent that E galaxies can be considered pressure supported and isothermal our proposed interpretation of the FP is consistent with a hierarchical dissipationless merging scenario for the formation of E galaxies in which the mass is increased and the central phase-space density is decreased in each step. This conclusion is supported by recent independent N-body simulations of Capelato, de Carvalho & Carlberg (1995) who find FP-like relations and deviations from the $R^{1/4}$ law quite similar to those found in this paper from a merging hierarchy of galaxies with constant mass-to-light ratio.

Correlations between luminosity and other observed dynamical quantities may also be envisaged in view of the dichotomy (or, perhaps, gradual change with luminosity) discussed above. For example, additional stellar dynamical effects can be a changing velocity structure with luminosity, such as rotation (Prugniel & Simien 1994) or velocity anisotropy (Merritt 1988). In fact, Burkert (1994) has shown that all galaxies in his sample with anisotropic velocity distributions, $(v/\sigma)^* < -0.33$, have $-0.25 < \delta b < 0$. Assume for example that an increase in radial velocity anisotropy is correlated with luminosity. For an Osipkov–Merritt Jaffe model (Binney & Tremaine 1987) we find that the line-of-sight velocity dispersion within an aperture of $R_e/5$ or $R_e/10$ increases by a factor of 1.2 when going from an isotropic velocity distribution (faint galaxies) to a very radial velocity distribution (bright galaxies) with an anisotropy radius $r_a/r_h = 0.1$. Over 5 absolute magnitudes in luminosity this variation in the observed line-of-sight velocity dispersion would translate into a value of $\gamma$ less than 1 (cf. equation (10)) if the galaxies were assumed to have identical velocity distributions. For our example this corresponds to $\alpha = 0.09$.

## 6. Conclusion

In this paper we have used a simple model for the phase-space density distribution in E galaxies derived from statistical mechanics of violent relaxation (Hjorth & Madsen 1991) to give a possible explanation of the small deviations from the $R^{1/4}$ law found by Burkert (1993) in a sample of 63 galaxies. In this interpretation, the observed departures from the $R^{1/4}$ law are essentially the signatures of the isothermal sphere, $\rho \propto \exp(\Psi)$, cf. equation (5) and Lima-Neto (1994). This view is also consistent with the numerical simulations of Londrillo et al. (1991). Further progress along this line could be obtained by analyzing a larger sample of high-quality profiles (e.g., Caon et al. 1993) in a consistent manner that can be carried out also for a set of models. To bridge the entire interval from giant to dwarf ellipticals it will be necessary to consider an alternative parameter describing the light profiles since the iterative determination of $\delta b$ does not converge for faint galaxies.

Our model and the observed correlation between the deviation of the light profile from a pure $R^{1/4}$ law (as measured by $\delta b$, cf. equation (4)) leads to a relation between the central potential and the luminosity for E galaxies.

Following this interpretation we have shown that the Fundamental Plane (FP) for the manifold of elliptical galaxies can be understood as the effect of a slightly changing structure of elliptical galaxies while they keep $R^{1/4}$ law profiles with deviations of only 0.1 mag (cf. Fig. 2). Adopting a universal mass-to-light ratio, $\mathcal{M}/\mathcal{L}$, for all galaxies we have determined the FP "$\mathcal{M}/\mathcal{L}$" exponent (cf. equation (3)) $\alpha \approx 0.12$ from these deviations in agreement with independent observational determinations in the infrared. We have thus demonstrated that it is possible, using simple stellar dynamics, to get a FP relation and that an $\mathcal{M}/\mathcal{L}$ variation with luminosity is not needed on purely dynamical grounds. In this picture the FP arises as a result of the slightly broken homology between E galaxies. Recently, a similar interpretation has been suggested to be viable for the FP correlations found for the Galactic globular cluster system (Djorgovski 1994).

To test this view it would be interesting, if possible, to establish the FP using IR array photometry. Likewise, to test for possible evolutionary effects (dynamical or chemical) it is important to investigate the FP correlations of E galaxies at significant redshifts.

As an alternative to the usual core-fitting technique we have determined the central phase-space densities of elliptical galaxies using their detailed light profiles as an indicator of the depth of the central potential. This method has the advantage of being seeing independent. We find that the central phase-space densities are comparable to or less than those of spiral galaxies ($\lesssim 10^{-6} M_\odot \text{pc}^{-3} (\text{km s}^{-1})^{-3}$) and follow a relatively weak dependence of luminosity, $f_c \propto \mathcal{L}_B^{-1.5}$. This indicates that ellipticals as well as spirals can merge into ellipticals without much dissipation and can occupy the FP without having a special IMF or distribution of the dark-matter. We attribute the high central surface-brightness of E galaxies observed with HST to dissipational processes, e.g., the formation of embedded nuclear disks or AGN.

Finally, we point out that our results have been reached in several tempi: First, we have made the case that the deviations from the $R^{1/4}$ law and their correlation with luminosity are evidence for non-homology of elliptical galaxies. Second, we have argued that the central potentials of ellipticals can be determined from our model. Third, in our discussion of the FP and phase-space densities, we have assumed a constant mass-to-light ratio (case (ii)). This assumption constitutes the simplest case, but in reality one could have a combination of non-homology and a variation of mass-to-light ratio with luminosity.

JH acknowledges discussions with Inger Jørgensen and financial support from the Carlsberg Foundation. Andi Burkert kindly provided the data points upon which Figure 3 is based. This work was supported in part by the Danish Natural Science Research Council (SNF) and the Theoretical Astrophysics Center (TAC) sponsored by the Danish National Research Foundation.